\documentclass[aps,prl,twocolumn,email,superscriptaddress,reprint]{revtex4}
\usepackage{amssymb}
\usepackage{amsmath}
\usepackage{natbib}
\usepackage{graphicx}
\usepackage{bbold}
\usepackage{color}
\usepackage{soul}
\usepackage{pdfpages}

\begin{document}
    \title{Deep learning of ultrafast pulses with a multimode fiber}
    \author{Wen Xiong}
    \affiliation{Department of Applied Physics, Yale University, New Haven, Connecticut 06520, USA}
    \author{Brandon Redding}
    \affiliation{Department of Applied Physics, Yale University, New Haven, Connecticut 06520, USA}
    \author{Shai Gertler}
    \affiliation{Department of Applied Physics, Yale University, New Haven, Connecticut 06520, USA}
    \author{Yaron Bromberg}
    \affiliation{Department of Applied Physics, Yale University, New Haven, Connecticut 06520, USA}
    \author{Hemant Tagare}
    \affiliation{Department of Radiology and Imaging Science, Yale University, New Haven, Connecticut 06520, USA.}
    \affiliation{Department of Biomedical Engineering, Yale University, New Haven, Connecticut 06520, USA.}
    \author{Hui Cao} 
    \email{hui.cao@yale.edu}
    \affiliation{Department of Applied Physics, Yale University, New Haven, Connecticut 06520, USA}
    \date{\today}
    \begin{abstract}
    Characterizing ultrashort optical pulses has always been a critical but difficult task, which has a broad range of applications. We propose and demonstrate a self-referenced method of characterizing ultrafast pulses with a multimode fiber. The linear and nonlinear speckle patterns formed at the distal end of a multimode fiber are used to recover the spectral amplitude and phase of an unknown pulse. We deploy a deep learning algorithm for phase recovery. The diversity of spatial and spectral modes in a multimode fiber removes any ambiguity in the sign of the recovered spectral phase. Our technique allows for single-shot pulse characterization in a simple experimental setup. This work reveals the potential of multimode fibers as a versatile and multi-functional platform for optical sensing.

    \end{abstract}
    \maketitle
   	\section{Introduction}
    Multimode fibers (MMFs) provide diverse degrees of freedom in space, spectrum, polarization and time, enabling a wide range of applications beyond their traditional role in communication. By manipulating the spatial degrees of freedom, an MMF can operate as a diffraction-limited microscope \cite{Papadopoulos12,Caravaca-Aguirre13,Ploschner15,French18}, a high-resolution spectrometer \cite{Redding12,redding2013all,Redding14, redding2014noise}, a radio-frequency wave sensor \cite{VallyeOL16, Sefler18}, an optical pulse shaper \cite{Carpenter15, Xiong16, Xiong17, Ambichl17, xiong2019nc}, a reconfigurable waveplate \cite{2018_Xiong_LSA} and a tailorable nonlinear element \cite{Wright15, Florentin17, Tzang18, chekhovskoy2018demand}. Previously we demonstrated that the intensity pattern formed by the interference of guided modes at the output of an MMF could be used to recover the spectral amplitude of input light \cite{Redding12,redding2013all,Redding14, redding2014noise}. Recovering the spectral phase, however, is more challenging because different frequencies do not interfere on a linear detector in a time-integrated measurement.
    
    Here, we propose a nonlinear time-integrated measurement of transmitted light through an MMF to extract the spectral phase of an optical pulse. Two-photon absorption on an array of detectors produces a nonlinear speckle pattern. From the speckle pattern we can retrieve the relative phase of different spectral components of the pulse, because those components interfere in the two-photon absorption process. Nonlinear optical processes have been widely used to characterize ultrafast pulses in the absence of a reference pulse \cite{weber1967method, kane1993single, iaconis1998spectral, lozovoy2004multiphoton, nicholson1999full}. However, many of these self-referenced techniques cannot determine the sign of spectral phase or the direction of time. For example, autocorrelation is commonly used to estimate the pulse width, but it always produces a temporally symmetric trace. Our new technique can resolve the direction of time, because an MMF (with uncontrolled bending/twisting) does not keep the symmetry of temporal inversion with phase conjugation. Compared to other pulse measurement methods such as FROG \cite{kane1993single}, SPIDER \cite{iaconis1998spectral}, MIIPS \cite{lozovoy2004multiphoton} and PICASO \cite{nicholson1999full}, our scheme has a simple experimental setup without moving parts, and allows for single-shot measurement, which is of particular importance when measuring unstable pulse trains.  
    
    The complexity of our approach is concentrated on phase retrieval from the nonlinear speckle pattern. Taking advantages of the overwhelming advancements in machine learning and deep neural networks, we employ deep learning for phase retrieval. Artificial neural networks were utilized for phase retrieval in FROG measurements \cite{krumbugel1996direct} and were demonstrated to outperform other phase retrieval algorithms \cite{zahavy2018deep}. FROG measurements based on second-harmonic generation suffer from the ambiguity in the direction of time, which would cause instabilities in the training of neural networks, unless additional constraint was imposed on the pulse shape. Our MMF-based technique does not have such a problem. Previously, deep neural networks were employed for imaging through multimode fibers \cite{rahmani2018multimode, fan2019deep}, but a large amount of data were needed for training.  It is much easier and less expensive to generate data numerically than experimentally to train the neural networks. However, it is difficult to accurately model a realistic fiber with unknown refractive index fluctuations and micro-bending/twisting to produce high-quality numerical data for training. Here we use the experimentally measured transmission matrix of the multimode fiber to calculate two-photon speckle patterns for training purpose. This hybrid method makes it easy and straightforward to generate numerical data for the specific fiber in the experiment. Moreover, noise in the measurement is included in the training process, making the neural network robust compared to conventional phase retrieval methods. Finally we combine machine learning with compressive sensing by representing the spectral phases of commonly seen pulses in a sparse basis, greatly reducing the number of parameters that need to be retrieved by the neural network. 
    
    \section{Principal of Operation}
    \begin{figure*}[t]
    	\includegraphics[width=2\columnwidth,keepaspectratio,clip]{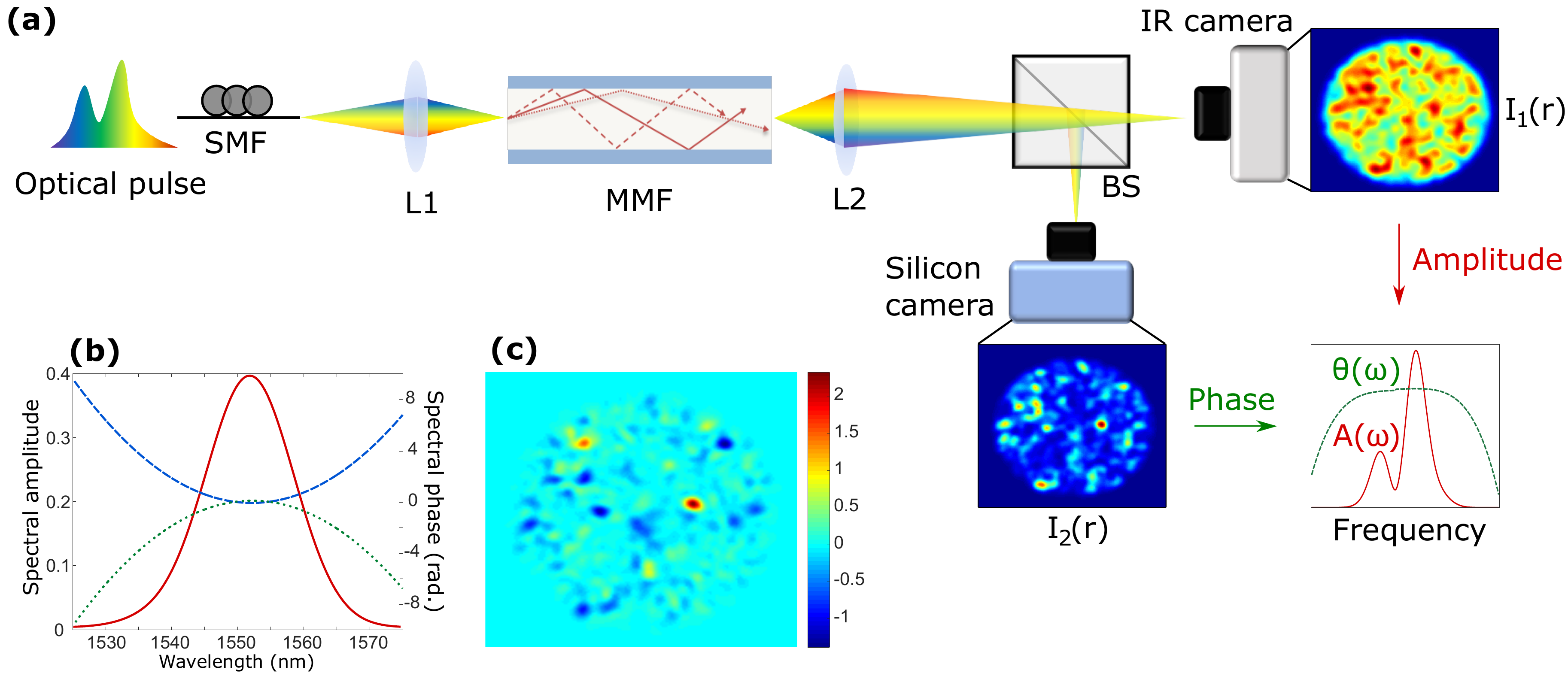}
    	\caption{(a) Experimental realization of the measurement scheme. At the input, an optical pulse at $\lambda$ = 1550 nm, delivered via a single mode fiber (SMF), is coupled to the calibrated MMF (Thorlabs FG105LCA). At the output, the IR camera (Xenics Xeva 1.7-640) records the time-integrated one-phone absorption pattern $I_1({\mathbf{r}})$. A silicon camera (Andor Newton DU940N-UV) detects the time-integrated two-photon absorption pattern $I_2(\mathbf{r})$. The amplitude $A(\omega)$ and phase $\theta(\omega)$ of the input pulse are recovered from $I_1(\mathbf{r})$ and $I_2(\mathbf{r})$ respectively. (b) Numerically simulated spectral amplitude $A(\omega)$ (red solid line), spectral phase $\theta(\omega)$ (blue dashed line), and the spectral phase with the flipped sign $-\theta(\omega)$ (green dotted line). (c) Difference in $I_2(\mathbf{r})$ between the pulse with the original and the flipped spectral phase.}
    	\label{fig:setup}
    \end{figure*}
    
    Our scheme relies on the speckle pattern formed at the end of a multimode fiber to provide a unique fingerprint of an optical pulse. Each speckle grain at the distal end provides a different sampling of the pulse. At the input, the pulse excites many guided modes with different propagation constants, and thus it experiences modal dispersion while propagating through the fiber. At the output, individual speckles are formed by different summations of all spectral components of the pulse, each with a varying amplitude and phase. The transmitted pulse displays distinct stretching and distortions from one speckle to another. The spectral amplitude of the input pulse is extracted from the time-integrated intensity measurement of output speckle pattern via one-photon absorption on the camera, as done previously in \cite{Redding12,Redding14}.  The spectral phase is recovered from the time-integrated nonlinear measurement of speckle pattern via two-photon absorption on a different camera. Since cameras detect all speckle grains in parallel, the amplitudes and phases of all spectral components of a pulse can be extracted with a single-shot measurement. An experimental realization of the proposed scheme is shown in Fig.~\ref{fig:setup}(a). 
    
    To use the speckle pattern as the fingerprint of a pulse, we first calibrate the spectral to spatial mapping of the MMF (step index, core diameter = 105 $\mu$m, numerical aperture = 0.22, length = 1.3 m). It requires a full-field measurement of the output light as a function of the input frequency. We use a frequency-tunable laser source, and the transmitted field is measured by off-axis holography in an interferometric setup \cite{Xiong16, 2018_Xiong_LSA}. The incident light is linearly polarized and one polarization of transmitted light is selected for detection. The complex field transmission coefficients measured at multiple frequencies $\omega$ are stored in a transmission matrix $T(\mathbf{r},\omega)$, where $\mathbf{r}$ denotes the spatial location at the fiber output. $T(\mathbf{r},\omega)$ relates the input spectral amplitude $A(\omega)$ and phase $\theta(\omega)$ to the complex output field $E_{\rm out}(\mathbf{r},t)$ for a fixed incident wavefront:  
    \begin{equation}
    E_{\rm out}(\mathbf{r},t) = \int T(\mathbf{r}, \omega) e^{-i\omega t} A(\omega)e^{i\theta(\omega)} d\omega.
    \end{equation}
    The time-integrated intensity pattern (linear speckle pattern) 
    \begin{equation}
    I_{\rm 1}(\mathbf{r}) = \int|E_{\rm out}(\mathbf{r},t)|^2dt 
    	= \int |T(\mathbf{r},\omega)|^2 |A(\omega)|^2 d\omega,
    	\label{eq:1p}
    \end{equation}
    is independent of the spectral phase $\theta(\omega)$. 
    
    The two-photon absorption pattern $I_{\rm 2}(\mathbf{r}) = \int |E_{\rm out}(\mathbf{r},t)|^4 dt$ can be expressed as  
     
 \begin{figure*}[t]
	\includegraphics[width=1.5\columnwidth,keepaspectratio,clip]{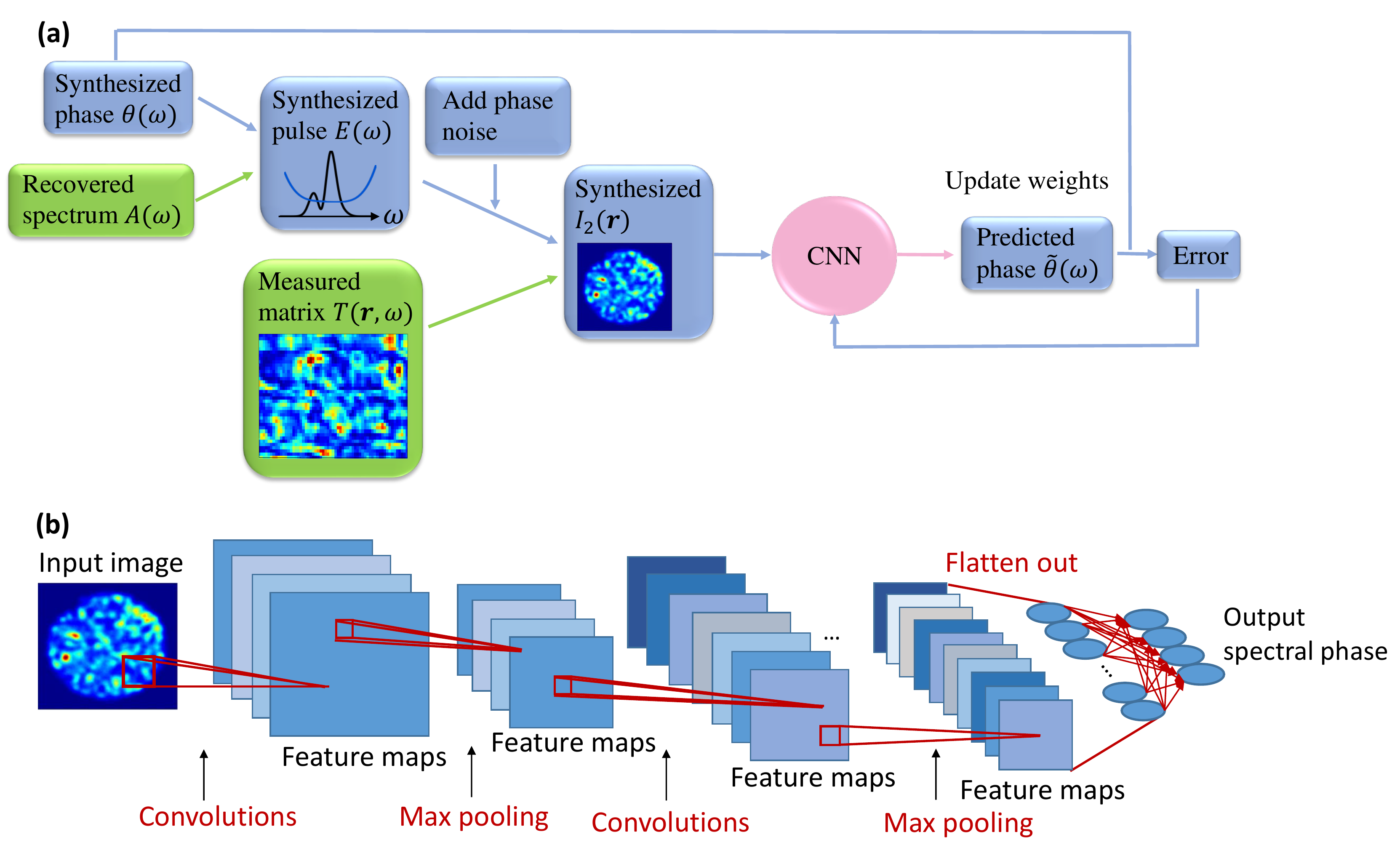}
	\caption{(a) Flowchart of the phase recovery algorithm. (b) Architecture of a convolutional neural network with convolutional layers, max pooling layers and fully-connected layers to retrieve spectral phase of a pulse from two-photon intensity pattern. }
	\label{fig:flow}
\end{figure*}

	\begin{widetext}
    \begin{equation}
    \begin{split}
    I_{\rm 2}(\mathbf{r}) 
    =\iiint d\omega_1 d\omega_2 d\omega_3 
    |T(\mathbf{r},\omega_1)| A(\omega_1)  |T(\mathbf{r},\omega_2)| A(\omega_2) |T(\mathbf{r},\omega_3)| A(\omega_3)  |T(\mathbf{r},\omega_1-\omega_2+\omega_3)| A(\omega_1-\omega_2+\omega_3) \\ e^{i[\theta(\omega_1)-\theta(\omega_2)+\theta(\omega_3)-\theta(\omega_1-\omega_2+\omega_3)]}
    e^{i[\phi(\mathbf{r},\omega_1)-\phi(\mathbf{r},\omega_2)+\phi(\mathbf{r},\omega_3)-\phi(\mathbf{r},\omega_1-\omega_2+\omega_3)]},
    \end{split} 
            \label{eq:2p}
    \end{equation}
	\end{widetext} 
	where $\phi(\mathbf{r},\omega)$ denotes the phase of $T(\mathbf{r},\omega)$. The dependence of $I_{\rm 2}(\mathbf{r})$ on $\theta(\omega)$ can be used to retrieve the spectral phase of input pulse. 
	
	When a random superposition of fiber modes is excited, due to the spatial complexity, the transmission matrix $T(\mathbf{r},\omega)$ possess no symmetry. The phases $\phi(\mathbf{r},\omega)$ of its elements  are randomly distributed over $(-\pi, \pi)$. If the input pulse is temporally reversed and phase conjugated, the sign of $\theta(\omega)$ is flipped. Since $\phi(\mathbf{r},\omega)$ remains the same, the phase of the transmitted field $\theta(\omega) + \phi(\mathbf{r},\omega)$ changes. Consequently, the two-photon speckle pattern $I_{\rm 2}(\mathbf{r})$ is modified. It is instructive to consider the complementary time domain picture, where due to the complex dynamics in the fiber, the temporal impulse response at each output position r is non-symmetric. Hence two time-reversed inputs, $E_{\rm in}(t)$ and $E_{\rm in}(-t)$ will yield two different temporal dynamics at the output, resulting in two different speckle patterns $I_2(r)$. With the experimentally measured transmission matrix, we calculate the two-photon pattern $I_{\rm 2}(\mathbf{r})$ for the synthesized amplitude and phase in Fig.~\ref{fig:setup}(b).  Figure~\ref{fig:setup}(c) presents the change in $I_{\rm 2}(\mathbf{r})$ when the spectral phase of a simulated pulse has its sign flipped. The relative change $\langle|\Delta I_{\rm 2}|\rangle/\langle I_{\rm 2}\rangle$, averaged over $\mathbf{r}$, is 0.13. Hence, the two-photon absorption pattern can eliminate the ambiguity with respect to temporal inversion with phase conjugation.
   
   \section{Deep learning}
   
   As shown in Eq.~(\ref{eq:2p}), the mapping from the spectral phase $\theta(\omega)$ of input pulse to the two-photon speckle pattern $I_{\rm 2}(\mathbf{r})$ at the fiber output is nonlinear and complex. It is very difficult to recover $\theta(\omega)$ from the measured $I_{\rm 2}(\mathbf{r})$. Conventional phase retrieval algorithms are sensitive to noise in the measurement, and thus cannot provide a reliable recovery. However, once the fiber transmission matrix is known, it is straightforward to calculate the output speckle pattern for any input pulses with Eq.~(\ref{eq:2p}). We deploy a convolutional neural network (CNN) to learn the inverse mapping from the output two-photon pattern to the input spectral phase. With noise incorporated to the network training, the CNN outperforms the standard phase retrieval algorithms \cite{zahavy2018deep}.
    
   Figure~\ref{fig:flow}(a) is the flowchart of the pulse recovery algorithm. First the amplitude spectrum $A(\omega)$ of the pulse is retrieved from the linear speckle pattern $I_{\rm 1}(\mathbf{r})$ with Eq.~(\ref{eq:1p}) \cite{Redding12,redding2013all}. Then we calculate $I_{\rm 2}(\mathbf{r})$ for various $\theta(\omega)$ using Eq.~(\ref{eq:2p}) with the calibrated $T(\mathbf{r},\omega)$ and the recovered $A(\omega)$. With these numerical data, we train the CNN for the specific spectrum of the probe pulse. The simulated nonlinear speckle pattern is the input of the CNN, and the predicted spectral phase $\tilde{\theta}(\omega)$ is compared to the known phase. Their difference is taken as the MAE (mean absolute error) defined as  $\overline{|\tilde{\theta}(\omega) - \theta(\omega)|}$. The error is propagated back through the network to update the weights in the CNN. After the training, a pulse with the same amplitude but an unknown phase is launched into the MMF, and the experimentally measured $I_{\rm 2}(\mathbf{r})$ is sent to the CNN to predict $\theta(\omega)$. The temporal field $E(t)$ of the pulse is finally obtained by applying a Fourier transform to the recovered spectral field $ E(\omega) = A(\omega) e^{i \theta(\omega)}$. 
   
   The basic architecture of a general CNN we adopted is shown in Fig.~\ref{fig:flow}(b). It extracts features of the speckle pattern by convolving it with spatial filters. Many filters are applied to the pattern to obtain an array of feature maps. The most important features are kept by a max-pooling layer and passed to the next convolutional layer. The last few layers of the CNN flatten out all the extracted features and map them to the desired output, i.e., the spectral phase. Specifically, we employ the architecture of Res-Net 18 \cite{he2016deep} in Pytorch machine learning library \cite{collobert2002torch}. Res-Net 18 is a small CNN architecture with less parameters to avoid overfitting and can be trained faster. The weights of the neural network are optimized using Adam \cite{kingma2014adam} for 1000 epochs with the initial learning rate set to $ 1\times$$10^{-4}$. The learning rate is reduced by a factor of 10 subsequently after 200, 400 and 800 epochs.

   To reduce the number of parameters that the CNN needs to predict, we represent the spectral phases in a sparse basis. For a chirped pulse, its spectral phase can be expressed as a polynomial, $\theta(\omega) = \sum_{i}\alpha_i (\omega - \omega_0)^i$, where $\omega_0$ is the central frequency of the pulse spectrum. The zeroth-order term $\alpha_0$ is a constant phase, which can be set to 0. The first-order term $i=1$ represents a linear phase chirp. $\alpha_1$ determines the time delay of the pulse, but does not affect the pulse shape or the two-photon pattern, so we set  $\alpha_1 = 0$. We keep the second-, third- and forth-order terms $i = 2, 3, 4$, which represent quadratic, cubic, quartic phase chirps. The higher order terms $i \geq 5$ are usually negligible, so we set $\alpha_{i \geq 5} = 0$. 

 If the signal consists of multiple pulses, the interference of these pulses in the spectral domain produces oscillations. The spectral phase exhibits a discontinuity at every local minimum of the amplitude spectrum. We therefore introduce a phase jump $\beta_j$ at the frequency $\omega_j$ corresponding to the $j$-th local minimum of $A(\omega)$,
\begin{equation}
\theta(\omega) = \sum_{i=2,3,4} \alpha_i (\omega - \omega_0)^i + \sum_{j} \beta_j\Theta(\omega_j)
\label{eq:phase}
\end{equation}  
$\Theta(\omega_j)$ is the Heaviside function with the discontinuity at frequency $\omega_j$. The magnitude of phase jump $\beta_j$ is within $(-\pi, \pi)$. With the parametrized spectral phase, the CNN only needs to predict the coefficients $\alpha_i$ and $\beta_j$ in Eq.~(\ref{eq:phase}). 

We numerically generate 10,000 pairs of spectral phases and two-photon patterns, 8000 of which are used for training and the rest for validation. The training takes about 8 hours on a 8-GPU AWS cluster. Once the CNN is trained, recovering the spectral phase from an experimentally measured two-photon pattern takes only a few seconds.
   
   \section{Noise suppression}

    The major difficulty for phase retrieval is the noise in the measurement. Experimentally there are two main sources of noise: the fiber instability and the camera noise. The integration time of the InGaAs camera and the silicon camera is adjusted when recording the linear and nonlinear speckle patterns, so that the signal to noise ratio (SNR) exceeds 100 for $I_{\rm 1}(\mathbf{r})$ and 50 for $I_{\rm 2}(\mathbf{r})$. With this SNR, the camera noise is negligible. The dominant noise comes from the fiber instability. Since the fiber is not thermally stabilized or mechanically isolated in our experiment, ambient temperature drift and/or external vibrations cause changes in the fiber refractive index. Consequently, the phase of transmitted light changes, and such change varies from one frequency to another. This means the fiber transmission matrix during the recording of speckle patterns for unknown pulses differs from the calibrated one. Such difference causes the failure of conventional phase retrieval algorithms. 
    
    To account for the fiber instability, we incorporate noise to the synthesized data during the training of the CNN. To evaluate this method, we measure the field transmission matrix of the same fiber twice. With the first transmission matrix (TM1), we generate 10,000 pairs of spectral phases and two-photon patterns to train the CNN. Using the second transmission matrix (TM2), we calculate the two-photon patterns with the spectral phases that have never been seen by the CNN. This set of data is used to test the trained CNN. By using two measured transmission matrices, we account for fiber instability in time. Typically the standard deviation of phase difference between the two matrices is about 0.2.   
   
   \begin{figure}
   	\includegraphics[width=\columnwidth,keepaspectratio,clip]{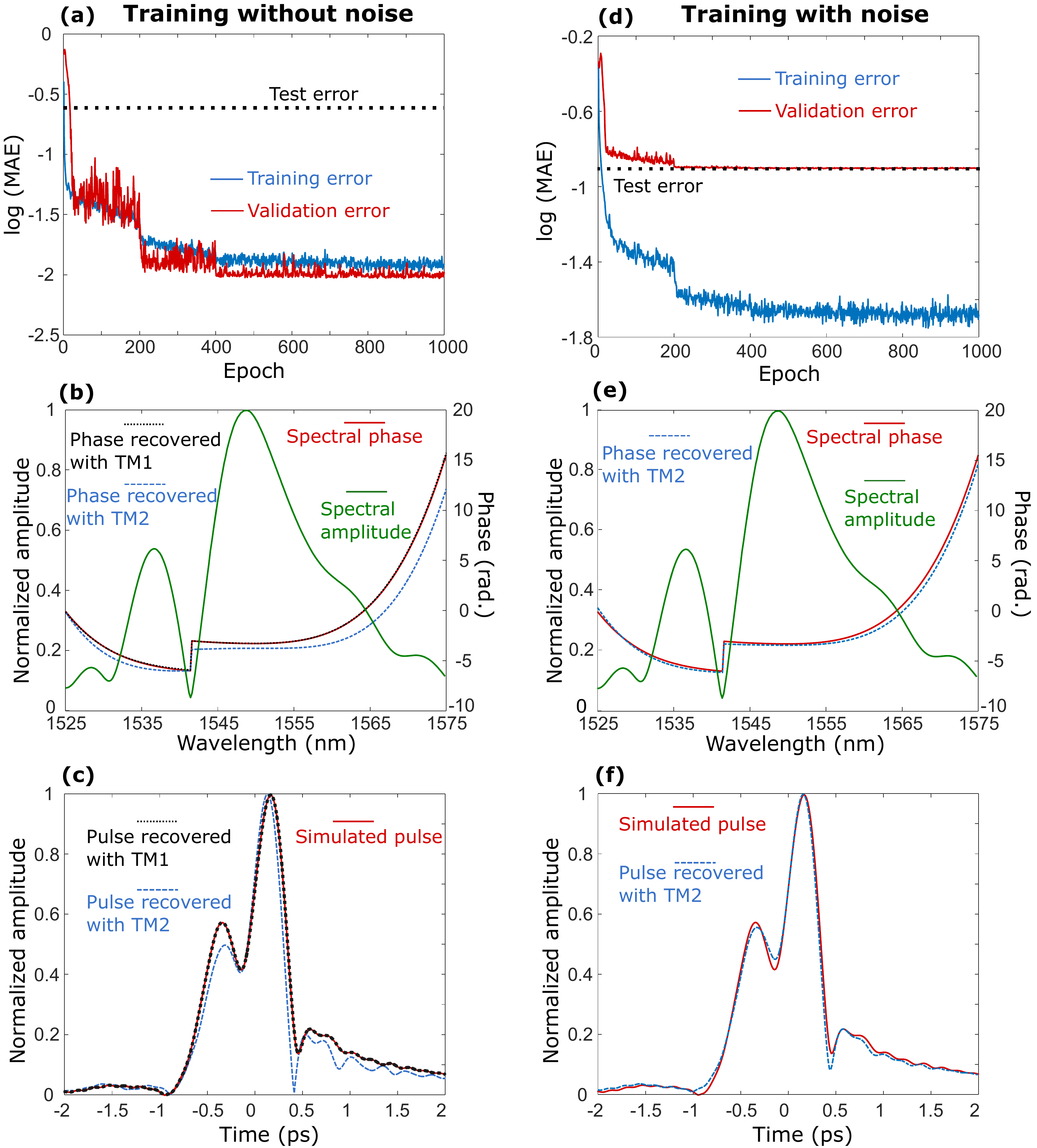}
   	\caption {Training and testing of convolutional neural network (CNN) with two measured fiber transmission matrices (TM1, TM2). (a,d) Error curve of CNN in the training process using TM1 without noise (a), TM1 with noise (d). (b,e) Spectral amplitude (green solid line) and phase (red solid line) of a pulse used for testing the CNN trained by TM1. The spectral phase recovered with the two-photon pattern generated by TM1 (black dotted curve) agrees well to the ground truth (red solid line), but the phase recovered with the two-photon pattern generated by TM2 (blue dashed curve) deviates from the ground truth (b). Incorporating noise to the training of CNN significantly reduces the deviation (e). (c,f) Temporal field amplitude of the pulse obtained from Fourier transform of the spectral amplitude and phase in (b,e). Deviation of the recovered temporal pulse shape (blue dashed line) from the ground truth (red solid line) is notably smaller using the CNN trained with noise (f) than that without noise (c).}
   	\label{fig:simulation_no_noise}
   \end{figure}

   \begin{figure*}
	\includegraphics[width=2\columnwidth,keepaspectratio,clip]{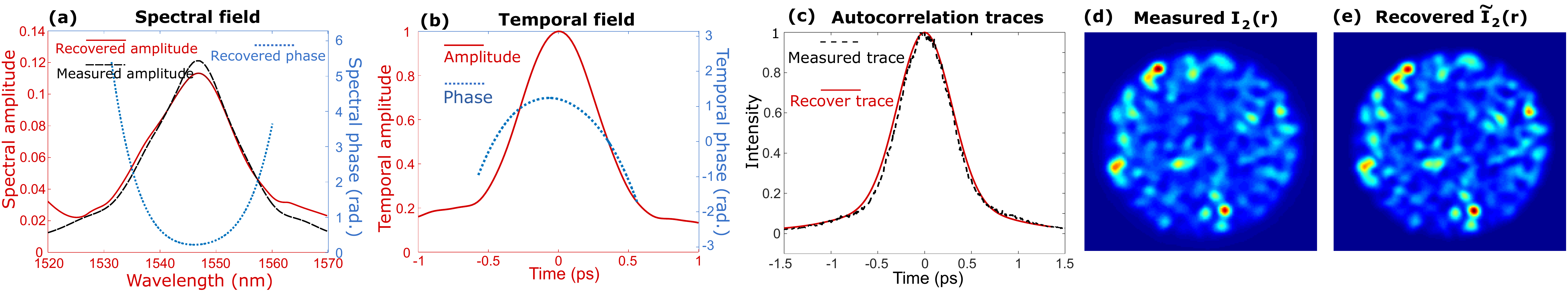}
	\caption{Recovery of a pulse propagating through the single-mode fiber with weak nonlinearity. (a) Recovered spectral amplitude (red solid line) and phase (blue dotted line) of the pulse, compared to the spectral amplitude measured by the optical spectrum analyzer (black dashed line). (b) Recovered temporal amplitude (red solid line) and phase (blue dotted line) of the pulse. (c) Measured temporal autocorrelation trace of the pulse (black dashed line) in good agreement with the autocorrelation trace of the recovered pulse (red solid line). (d) Experimentally measured and (e) recovered two-photon speckle patterns match well. }
	\label{fig:pulse1}
	\end{figure*}
   
   Without accounting for the phase fluctuations in the training of CNN, its prediction has limited accuracy. To illustrate this, we consider a pulse with the amplitude spectrum shown in  Fig.~\ref{fig:simulation_no_noise}(b) (green line). The spectral phase $\theta(\omega)$ is generated by Eq.~(\ref{eq:phase}), with only one phase jump $\beta_1$ at 1541.4 nm (red curve). 10,000 phase spectra are synthesized in the range of $\alpha_2 \in (-0.3, 0.3)$, $\alpha_3 \in (-0.05, 0.05)$ and $\alpha_4 \in (-0.005, 0.005)$, $\beta_1 \in (-\pi, \pi)$. 10,000 two-photon absorption patterns $I_{\rm 2}(\mathbf{r})$ are calculated with TM1: 8,000 for training the CNN and 2,000 for validation. Training and validation errors are obtained from the corresponding data sets.  Figure~\ref{fig:simulation_no_noise}(a) shows the training and validation error during the 1000 epoch training process. The training error is similar to the validation error. The minimum validation error is 0.01. A sample of the recovered spectral phase and the temporal pulse shape in the validation set is shown in Fig.~\ref{fig:simulation_no_noise}(b)-(c).  Since the CNN is trained by TM1, the prediction of $\theta(\omega)$ (black dotted curve) from $I_{\rm 2}(\mathbf{r})$ generated by TM1 (validation data) is accurate. For the 2,000 validation spectra, $I_{\rm 2}(\mathbf{r})$ are also calculated with TM2 as the test date set. The prediction of $\theta(\omega)$ from $I_{\rm 2}(\mathbf{r})$ generated by TM2 (blue dashed curve) is less accurate. The test error is 0.25, as indicated by the black dotted line in Fig.~\ref{fig:simulation_no_noise}(a), significantly higher than the validation error of 0.01. Such increase of error mainly results from the fiber instability, captured in the difference between TM1 and TM2. 
   
   To take into account the phase fluctuations in the transmission matrix, we add a random phase noise to the transmission matrix when generating the training data for the CNN. The random phase varies from column to column, each column corresponding to one frequency. It simulates the phase difference between TM1 and TM2, which changes with frequency. We incorporate different random phases in TM1 to compute the two-photon pattern for every synthesized spectral phase. We tune the standard deviation of the phase noise to minimize the error of CNN. With a standard deviation of 0.47, which is close to the phase fluctuations of the measured transmission matrix, we obtain the smallest error for $\theta(\omega)$ recovered from $I_{\rm 2}(\mathbf{r})$ generated by TM2 (with CNN trained with $I_{\rm 2}(\mathbf{r})$ generated by TM1). Compared to the case without noise, the test error is reduced from 0.25 to 0.14, as can be seen from the shift of the black dotted line from Fig.~\ref{fig:simulation_no_noise}(a) to (b). The validation error (from $I_{\rm 2}(\mathbf{r})$ generated by TM1) increases to 0.13, similar to the test error but significantly larger than the training error of 0.02. The gap between the training error and the validation error is the effect of overfitting, because the phase noise is not a feature that can be learned by the CNN. Instead the CNN is trained to ignore the noise. After the training with noise, the recovered spectral phase and temporal pulse shape in Fig.~\ref{fig:simulation_no_noise}(e,f) agree well with the synthesized ones.  

   \section{Experimental demonstration}
   
	   In our measurement scheme, the nonlinear process (two-photon absorption) occurs at the detector, not in the fiber. The propagation of optical pulse in the multimode fiber must be linear, otherwise the speckle pattern at the fiber output would vary with the incident pulse energy, making it extremely difficult to extract the temporal pulse shape. Fortunately optical nonlinearity is weak in a MMF, as a large fiber core reduces the energy density. Furthermore, when many guided modes are excited in the fiber, modal dispersion stretches the pulse temporally, lowering the peak power. 
	   
	   In contrast, optical nonlinearity in a single mode fiber (SMF) can be significant for short pulses. We use our new scheme to characterize the femtosecond laser pulses transmitted through an one-meter-long SMF. By varying the pulse energy coupled into the SMF, we can tune the strength of optical nonlinearity, which will distort the pulse shape.  
	   
     \begin{figure}
   	\includegraphics[width=1\columnwidth,keepaspectratio,clip]{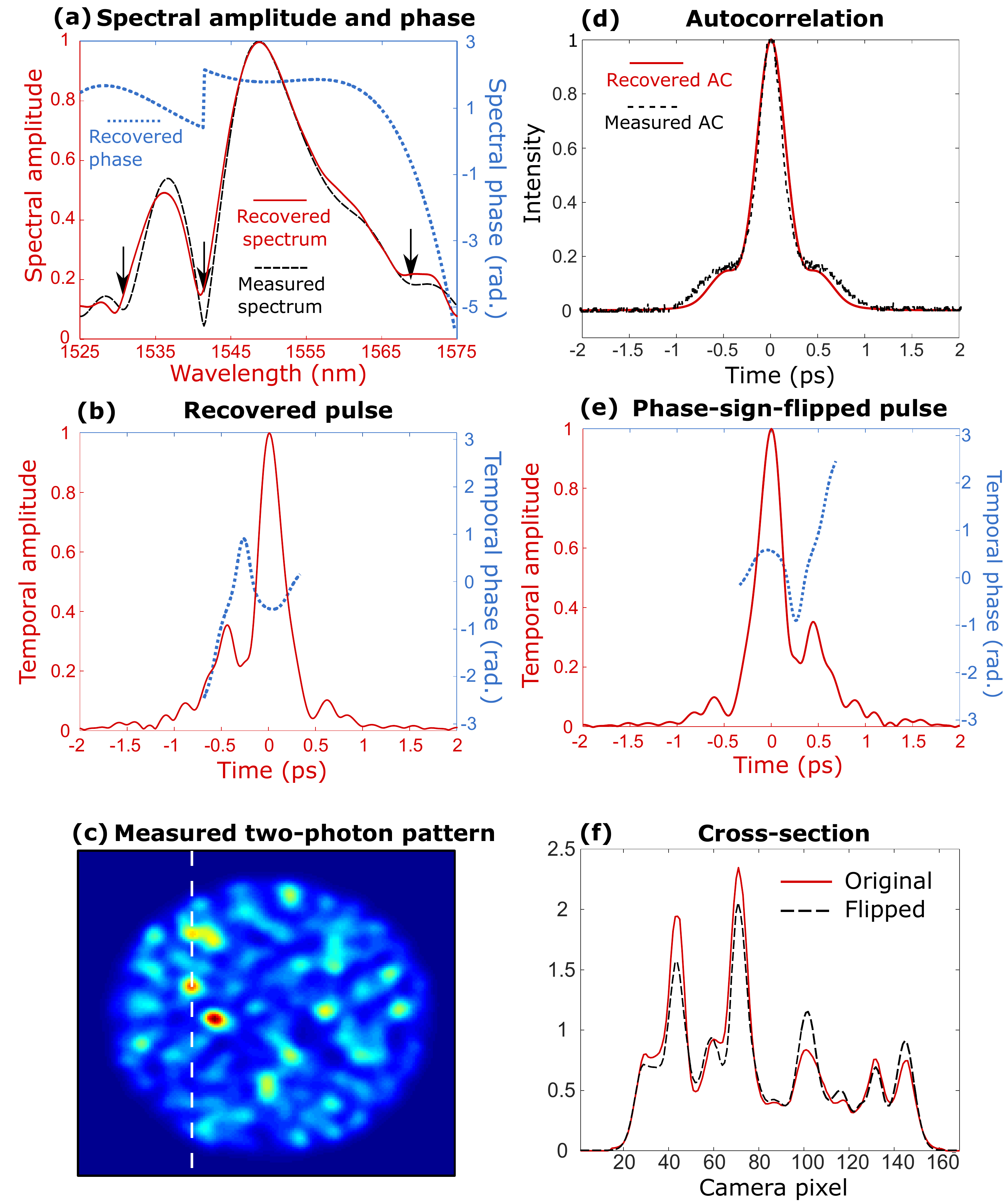}
   	\caption{Recovery of a pulse propagating through the single mode fiber with strong nonlinearity. (a) Recovered spectral amplitude (black dashed line) and phase (blue dotted line) of the pulse, compared to the spectral amplitude measured by the optical spectrum analyzer (red solid line). (b) Temporal amplitude (red solid line) and phase (blue dotted line) of the recovered pulse. (c) Measured two-photon pattern of the pulse transmitted through the MMF. (d) Experimentally measured temporal autocorrelation trace of the pulse (black dashed line) in good agreement with the autocorrelation trace of the recovered pulse (red solid line). (e) Temporal amplitude (red solid line) and phase (blue dotted line) of the pulse with the sign of the spectral phase flipped. (f) Two-photon intensities along the cross-section marked by the white dashed line in (c) of the recovered pulse and its time-reversed copy.}
   	\label{fig:pulse3}
   \end{figure}
   
   We first test the case of very weak nonlinearity. Nearly transform-limited pulses from a femtosecond laser (NKT, Onefive Origami) at 1550 nm are sent through the SMF. The transmitted pulses are then launched into the MMF. The spectral amplitude of the pulse is reconstructed from the linear speckle pattern at the MMF output. As shown in Fig.~\ref{fig:pulse1}(a), it agrees well with that measured by an optical spectrum analyzer, validating the accuracy of the spectral amplitude recovery. The spectral phase are parameterized according to Eq.~(\ref{eq:phase}). The phase discontinuities vanish, since the amplitude spectrum displays no local minimum. The three phase chirp terms ($i = 2, 3, 4$) are in the range specified in the last section. The recovered spectral phase is nearly constant across the frequency range of significant spectral amplitude. From the recovered spectral amplitude and phase, we reconstruct the temporal amplitude and phase of the pulse, as plotted in Fig.~\ref{fig:pulse1}(b). 
   
   To quantitatively estimate the accuracy of reconstruction, we measure the pulse transmitted through the SMF with an autocorrelator (Femtochrome FR-103XL). In spite of its ambiguity in recovery of a pulse shape, the autocorrelation trace is a useful metric to evaluate the accuracy of the pulse shape retrieved with other methods. We calculate the autocorrelation trace of the recovered pulse and it agrees well with the measured one in Fig.~\ref{fig:pulse1}(c). Using the recovered spectral amplitude and phase, we calculate the nonlinear speckle pattern $\tilde{I}_{\rm 2}(\mathbf{r})$ in Fig.~\ref{fig:pulse1}(d). It bears striking similarities to the experimentally measured one $I_{\rm 2}(\mathbf{r})$ in Fig.~\ref{fig:pulse1}(d). Both patterns are normalized. Their difference, given by the standard deviation  $\epsilon = \sqrt{ \int |I_{\rm 2}(\mathbf{r}) - \tilde{I}_{\rm 2}(\mathbf{r}) |^2 d\mathbf{r} }$, is 0.11. The excellent agreements obtained for both the autocorrelation traces and the nonlinear speckle patterns validate the recovered pulse shape.

   By coupling more power into the SMF, we enhance the nonlinear processes such as self-phase modulation, cross-phase modulation and four-wave mixing, which cause spectral and temporal distortions of the pulse. The amplitude spectrum of the pulse transmitted through the SMF is reconstructed with the linear speckle pattern, and it agrees well with the measurement by the OSA in Fig.~\ref{fig:pulse3}(a). The spectrum is severely distorted from that in  Fig.~\ref{fig:pulse1}(a), and it features three local minima (marked by arrows). When synthesizing the spectral phase, we only consider the phase discontinuity at $\lambda$ $\simeq$ 1541.4 nm, because the other two local minima at 1530.6 nm and 1568.8 nm are too shallow to affect the temporal pulse shape. The recovered spectral phase, plotted by the blue dotted line in Fig.~\ref{fig:pulse3}(a), exhibits a phase jump of 1.75 at $\lambda$ = 1541.4 nm. The temporal amplitude and phase of the pulse are then obtained by Fourier transform of the recovered spectral field and plotted in Fig.~\ref{fig:pulse3}(b). The pulse is asymmetric and has a side lobe. The autocorrelation trace for the recovered pulse has a good agreement with that measured by the autocorrelator in Fig.~\ref{fig:pulse3}(d). 
   
   By changing the sign of the spectral phase, the temporal field is inversed. As seen in  Fig.~\ref{fig:pulse3}(e), the side lobe is moved from the front to the tail of the main pulse. While the autocorrelation trace remains the same, the nonlinear speckle pattern in Fig.~\ref{fig:pulse3}(c) changes. In Fig.~\ref{fig:pulse3}(f), we plot the intensity over a cross-section of the pattern [white dashed line in Fig.~\ref{fig:pulse3}(c)] for the recovered pulse and the time-inversed pulse. They display significant differences, allowing the CNN to differentiate between the pulse and its time-inversed copy.

   \section{Discussion and conclusion}
   
   In summary, we demonstrate a novel method of characterizing spectral phases of ultrafast pulses with a multimode fiber (MMF). The propagation of the pulse in the MMF remains linear, and is calibrated by a field transmission matrix. The nonlinear process (two photon absorption) at the MMF output induces interference of different spectral components in the pulse, thus the nonlinear speckle pattern encodes the spectral phase. The complex interference eliminates the ambiguity in the sign of spectral phase, allowing the direction of time to be recovered. The spectral phases are retrieved by a deep neural network, which is trained with the data numerically synthesized with the experimentally measured fiber transmission matrix and the spectral amplitude recovered from the linear (one-photon absorption) speckle pattern. We combine machine learning with compressive sensing by representing the spectral phase in a sparse basis to dramatically reduce the number of parameters that the neural network predicts. Experimental noise is incorporated in the training process, making the trained network robust again fiber instability.

 We think the most attractive feature of the multimode fiber based pulse characterization scheme presented here is its simplicity. It does not need a reference pulse \cite{Villiger17, xiong2019multimode}, allowing stand-alone characterization of ultrafast pulses. Experimentally, it requires only a commercially available multimode fiber and two cameras. The InGaAs camera records the linear speckle pattern for retrieval of spectral amplitude, and the Silicon camera records the nonlinear speckle pattern for spectral phase recovery. While the cost of multimode fibers and the silicon camera is low, the InGaAs camera is expensive and could therefore limit adoption of this technique. Since the nonlinear speckle pattern also encodes the information of spectral amplitude of the pulse, it may be used to recover the amplitude spectrum in addition to the phase spectrum. By training a deep neural network to perform an inverse mapping from the nonlinear speckle pattern to the spectral amplitude and phase, it may be possible to characterize a pulse using only the silicon camera, foregoing the need for an expensive InGaAs camera. Of course, the InGaAs camera is still needed for the calibration of the fiber transmission matrix, but this could potentially be performed at the factory, if the MMF is thermally and mechanically stabilized and spliced to a SMF to ensure repeatable coupling \cite{Redding14}.% In addition, recent developments in the characterization of MMFs indicate that mode propagation in MMFs can in some cases be predicted analytically \cite{Ploschner15}. This could potentially enable calibration-free MMF sensors or at least dramatically simplify the calibration process.
 
 By using a silicon camera to measure the 2-photon speckle pattern, the current approach is limited to characterizing pulses in the near infrared spectrum (below the bandgap of silicon). However, the general approach of using a multimode fiber and measuring the 2-photon speckle pattern could be extended to other spectral regions by first imaging the end of the multimode fiber onto a second-harmonic-generation material and then recording the speckle pattern formed at the second harmonic frequency. In addition, this scheme can be easily tuned to characterize pulses of varying length. Specifically, the temporal resolution and the temporal extent of an optical pulse which can be measured using the MMF technique presented here depends on the spectral resolution of the fiber and the bandwidth over which we calibrate the fiber. The spectral resolution of the MMF employed in the current experiment is 0.24 nm, thus the temporal range of measurement is 30 ps. The fiber transmission matrix is calibrated in the wavelength range of 50 nm, giving a temporal resolution of 160 fs. By calibrating the fiber over a larger bandwidth, one could measure temporally shorter pulses, while using a longer fiber with finer spectral resolution would enable the measurement of pulses extending over a longer period of time.

    \section{Acknowledgment}
    The authors acknowledge Siyuan Dong, Yunzhe Li and Lei Tian for fruitful discussions on deep neural networks. We thank Sebastian Popoff for his contribution to the early stage of this project. This work is supported by US National Science Foundation under the Grant Nos. ECCS-1509361 and ECCS-1809099. 
    
\bibliography{ref}

\end{document}